\documentclass[apj]{emulateapj}
\usepackage{epsfig}
\usepackage{changebar}
\usepackage{natbib}

\def\boldsymbol{\bf}

\bibpunct{(}{)}{;}{a}{}{,}

\slugcomment{To be submitted to ApJ}

\shorttitle{South Pole Atmosphere}
\shortauthors{Bussmann et al.}

\usepackage{amsmath}
\usepackage{amssymb}

\begin{document}

\title{Millimeter Wavelength Brightness Fluctuations of the Atmosphere Above the South Pole}

\author{R. S. Bussmann\altaffilmark{1,2}}
\affil{Department of Astronomy, University of Arizona, 
Tucson, AZ 85721}
\email{rsbussmann@as.arizona.edu}

\author{W. L. Holzapfel\altaffilmark{1}}
\affil{Department of Physics, 
University of California, Berkeley, CA 94720}
\email{swlh@cosmology.berkeley.edu}

\and

\author{C. L. Kuo\altaffilmark{3}}
\affil{Jet Propulsion Laboratory, Pasadena, CA 91125}
\email{clkuo@astro.caltech.edu}

\altaffiltext{1}{Department of Physics, 361 LeConte Hall, University of
California at Berkeley, Berkeley, CA 94720}
\altaffiltext{2}{Department of Astronomy, University of
California at Berkeley, Berkeley, CA 94720}
\altaffiltext{3}{Department of Physics, California Institute of Technology,
Pasadena, CA 91125}

\begin{abstract}
We report measurements of the millimeter wavelength brightness fluctuations 
produced by the atmosphere above the South Pole made with the Arcminute Cosmology
Bolometer Array Receiver ({\sc Acbar}).
The data span the 2002 Austral winter during which 
{\sc Acbar} was mounted on the Viper telescope at the South Pole.
We recover the atmospheric signal in the presence of instrument noise by
calculating the correlation between signals from distinct elements of the
{\sc Acbar} bolometer array.  
With this method, it is possible to measure atmospheric brightness fluctuations
with high SNR even under the most stable atmospheric conditions.
The observed atmospheric signal is characterized by the parameters of the 
Komolgorov-Taylor (KT) model, which are 
the amplitude and power law exponent describing the atmospheric power spectrum, 
and the two components of the 
wind angular velocity at the time of the observation.  
The KT model is typically a good description of the observed fluctuations, and fits to the
data produce values of the Komolgorov exponent that are consistent with
theoretical expectations. 
By combining the wind angular velocity results with measurements 
of the wind linear velocity, we find that the altitude of the observed 
atmospheric fluctuations is consistent with the distribution of water vapor 
determined from radiosonde data. 
For data corresponding to frequency passbands centered on 
150,~219,~and~274~GHz, we obtain median fluctuation power amplitudes of 
[10, 38, 74] mK$^2$ rad$^{-5/3}$ in Rayleigh-Jeans temperature units.  
Comparing with previous work, we find that these median amplitudes are approximately 
an order of magnitude smaller than those found at the South Pole during the 
Austral summer and at least 30 times lower than found at the ALMA site in the 
Atacama desert.   

\end{abstract}


\section{Introduction}\label{sec:intro}

The geographic South Pole is one of the most important sites for astronomical
observations at mm and sub-mm wavelengths. 
This is primarily due to the relatively high altitude and the lack of water 
vapor in the atmosphere.
In addition, the stability of the atmosphere makes it especially well suited
for observations of the cosmic microwave background (CMB) radiation.
The recent detection of CMB polarization by DASI \citep{kovac02}
and fine scale CMB anisotropy measurements by {\sc Acbar}
\citep[hereafter K2004]{kuo04} precede a wave of new instruments to be 
deployed at the South Pole in the next few years.
Compared to
balloon-borne experiments, ground based experiments can make use of larger
mirrors and achieve longer continuous observation time.
It was shown in \citet{runyan03} that the background limited photon 
noise of the {\sc Acbar} experiment is only 50\% more than that 
achieved in the balloon-borne BOOMERanG experiment \citep{crill03}. 
For many types of observations at the South Pole, atmospheric brightness 
fluctuations will not significantly increase the noise above the background 
limit. 

The impact of atmospheric fluctuations on
astronomical observations at mm wavelengths has been studied by 
several authors. Most of this work is concerned with distortions
of the wavefront as the radiation propagates through the 
atmosphere and the effects this has on high resolution 
interferometry \citep[e.g.][]{lay97}. 
Previous studies of atmospheric brightness fluctuations at mm and cm-wavelengths
have been limited by sensitivity \citep[][hereafter LH2000]{wollack97,lay00}. 
Furthermore, this is the first characterization of the atmospheric brightness 
fluctuations made during the Austral winter at the South Pole.  
It is expected that the fluctuation power will be lower than during the Austral 
summer due to lower ambient temperatures and lower precipitable water vapor.

We fit the observed atmospheric induced correlations between detectors in the 
{\sc Acbar} data with a Kolmogorov-Taylor (KT) model for the angular power spectrum
of atmospheric fluctuations.   
The high SNR with which we measure the atmospheric
fluctuations makes it possible to determine the amplitude and exponent 
of the KT power law and wind angular velocity for our atmospheric model. 
Combined with 
radiosonde measurements of wind linear velocity, we can infer the altitude at which the
fluctuation signals arise.
This work provides a detailed characterization of fluctuations in the atmosphere
above the South Pole that can be used to compare sites and predict the performance
of instruments and observation strategies. 

In Section~\ref{sec:genprop}, we 
give a general overview of the environment of the South Pole as pertains to
millimeter wavelength astronomical observations.
Section~\ref{sec:model} describes the spatial 
(Kolmogorov) and temporal (Taylor) components of the model.  In 
Section~\ref{sec:observations}, we detail the instrument specifications and 
explain the observing strategy.  Sections~\ref{sec:data}~and~\ref{sec:modfit} 
discuss the methodology used to process the data and compare to models for
the atmospheric fluctuations.
In Section~\ref{sec:results}, we present the 
results of this analysis, including the exponent of the power law, 
the amplitude of the fluctuations, measurements of the wind angular 
velocity, and the scale height of the fluctuations.  
Finally, in Section~\ref{sec:conclusions} we make some 
concluding remarks, including a comparison of our results with those 
obtained at the South Pole during the Austral summer and at the
ALMA site in Chile. 

\section{Atmosphere above South Pole}\label{sec:atmodel}

\subsection{General Properties}\label{sec:genprop}
The large scale air flow
at the South Pole is dominated by a katabatic (moving downward due to 
cooling) wind from the Antarctic plateau driven toward the coast. 
At the South Pole Station, radiosonde meteorology measurements are taken 
twice a day during the summer, and once a day in winter.
These data, which include the pressure, temperature, 
relative humidity, and wind angular velocity
as a function of altitude, provide important 
information on the atmospheric properties. 
The weather data show that the surface wind direction 
is relatively constant, with higher wind speed in the winter than in the summer.
It is also found that during the winter months, 
the cold surface creates a layer of 
temperature inversion where the air temperature rises from
$\sim -60^\circ$C at ice level to $\sim -40^\circ$C at an altitude of
approximately $200\,$m.

Using balloon-borne microthermal probes, \citet{marks99} found 
this temperature inversion layer to be turbulent.
The AASTO team also confirmed this result with a non-intrusive SODAR 
system \citep{aasto03}. The lack of radiative heating during the winter and the 
featureless landscape near the South Pole imply that the
turbulence in the inversion layer is due to shear-induced instability. 
The high correlation between the horizontal wind speed and temperature 
fluctuation reported by AASTO also supports this point of view.
The low-altitude atmospheric turbulence creates variations in the refraction index, 
and is of 
primary concern for optical seeing \citep{marks02,aasto03}. 

Observations at  mm wavelengths are limited by the emission 
of radiation by molecules in the atmosphere.
The primary components of the air (O$_2$ and N$_2$) are very well mixed; 
these ``dry'' components produce a uniform background of radiation that 
contributes photon noise to the detector. 
Water vapor in the atmosphere not only radiates
at mm wavelengths, but also fluctuates in its mass fraction, 
resulting in brightness temperature fluctuations. 
Because of the low temperature immediately above the ground in the winter, 
the integrated water 
vapor pressure is small in the turbulent boundary layer ($<$200m). 

\begin{figure}[!bp]
\begin{center}
\leavevmode
\epsscale{1.0}
\plotone{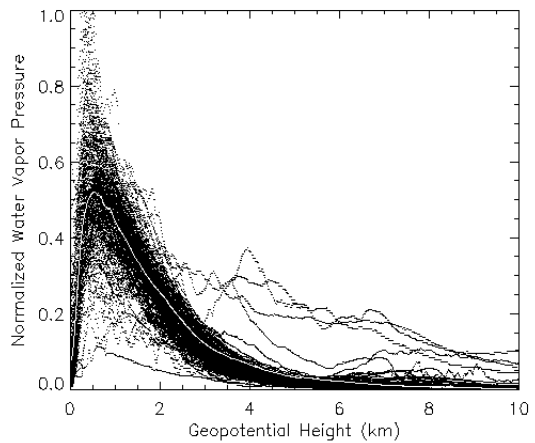}
\end{center}
\caption{Water vapor pressure as a function of geopotential height above ice level at 
the South Pole as measured by radiosonde data, and normalized to the maximum
observed value.  Black lines represent water vapor pressure 
profiles for each day during the Austral winter; the white line is the average of all the 
profiles.  Most of the water vapor pressure originates in a broad layer in 
the atmosphere between altitudes of 0.3 and $2\,$km.  Anomalous profiles well
above or below the average appear to be due to measurement errors.\label{p_wv}}
\end{figure}

In the optically thin limit, the water vapor contribution to sky brightness temperature 
is a function of the optical depth $\tau$ and thermodynamic temperature 
$T$; in other words, it is proportional to the line of sight integral of water vapor 
pressure. Approximating the 
saturated water vapor pressure as an exponential function in $T$ (following 
the Goff-Gratch equation\footnote{http://cires.colorado.edu/\~{}voemel/vp.html}),
the water vapor pressure profile can be derived from the publicly available 
archived radiosonde data.\footnote{ftp://ice.ssec.wisc.edu/pub/raob/}
Fig.~\ref{p_wv} shows the water vapor profiles measured with 
radiosonde data during 2002. 
Based on this data, we expect that 
the microwave radiation from water vapor at the South Pole 
comes predominantly from a thick layer between 0.3~km and 2~km 
above the ice. 

\begin{figure}[!tp]
\begin{center}
\leavevmode
\epsscale{1.0}
\plotone{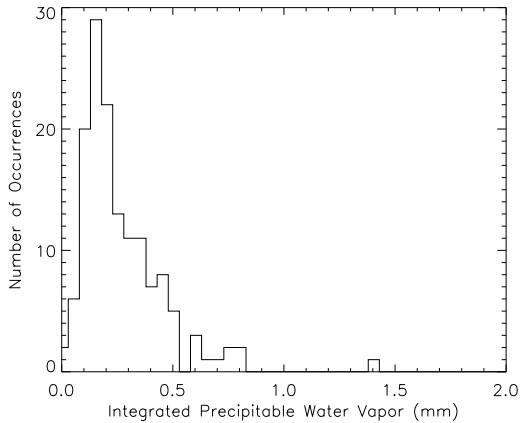}
\end{center}
\caption{Distribution of integrated precipitable water vapor (in mm) at the South Pole 
for the 2002 Austral winter.  The mean value of $0.26\,$mm is extremely small 
when compared to other well characterized astronomical sites.\label{fig:pwvhist}}
\end{figure}

One common way to characterize the quality of an observing site is to report the 
distribution of the integrated precipitable water vapor for an extended period 
of time.  Previous studies of sky opacity in the sub-mm produced median precipitable water 
vapor levels at the South Pole of about 0.25~mm \citep{chambal94}.  The radiosonde data in 
Fig.~\ref{p_wv} can be integrated to determine the precipitable water vapor for each 
day.
Fig.~\ref{fig:pwvhist} shows a histogram of the precipitable water vapor measurements
over the 2002 Austral winter.  
The mean precipitable water vapor during winter is 0.26$\; \pm \; 0.2\,$mm.
This is remarkably low when compared to the best six months at other well characterized sub-mm sites such as 
Mauna Kea (1.65$\,$mm) and the Atacama desert (1.00$\,$mm)
\citep{lane98}.

\subsection{The Model}\label{sec:model}

The Kolmogorov Turbulence (KT) model describes the cascade of turbulent energy from the large scales
on which it is input, to the small scales where it is dissipated. 
Large scale turbulence in the atmosphere above the South Pole is produced by processes 
such as wind shear and convection. These processes produce fluctuations with scales 
that exceed or are comparable to the thickness of the turbulent layer.
Eventually, the turbulent energy is dissipated on scales comparable to $1\,$mm, which
are far smaller than we are capable of observing.
The fluctuation angular power 
spectrum between the scales of energy injection and viscous dissipation, commonly 
known as the inertial subrange, is described by a power law with an exponent of $-11/3$.
In the work described here, the scales being observed all lie in the inertial subrange 
and the fluctuation power should be well described by the KT model. 

We assume the atmospheric signal seen by the telescope is dominated
by fluctuations in a layer of turbulence with thickness $\Delta h$, 
located at height $h$ above the ground.
The telescope is pointed at an elevation angle $\epsilon$, and 
the observed patch of sky subtends an angle of $\Delta \theta \sim 3^{\circ}$.
Due to the relatively small size of the observed field and the typical
altitude of the atmospheric water vapor, 
\begin{equation}\label{eq:smallangle}
\Delta h \gg \frac{h}{\sin \epsilon} \Delta \theta\;,
\end{equation}
and it is safe to assume that we are observing the projection of
the full 3-D Kolmogorov turbulence.
In this limit, the thickness of the turbulent layer only affects the projected amplitude 
of the fluctuations. 
In addition, we are able to employ the small angle approximation $(\Delta \theta \ll 1)$
and avoid the complexity of spherical harmonic functions. 
Under these assumptions, the angular power spectrum follows the 
3-D isotropic KT power law ($P(k) \propto k^{-11/3}$), and can be parametrized as

\begin{equation}\label{eq:palpha}
P(\alpha)=B^2_{\nu}\ {\rm sin}(\epsilon)^{-8/3} (\alpha_x^2+\alpha_y^2)^{-b/2}\;,
\end{equation}

\noindent where $k$ is the spatial wave number, 
$\alpha=kh/\sin\epsilon$ is the angular wave number, and $B_{\nu}^2$ is the amplitude
fluctuation power at observing frequency $\nu$ (in GHz). 
From the KT model for atmospheric turbulence, we expect that $b=11/3$; however, initially, 
we leave it as a free parameter in the model fitting. 
This is the same model used by \citet{lay97} and LH2000
to characterize atmospheric brightness fluctuations. 
$B^2_{\nu}$ is the amplitude of the power spectrum normalized to 
observations at the zenith and has units of ${\rm mK}^2 \; {\rm rad}^{-5/3}$.
We provide a more rigorous derivation of this relationship in Appendix \ref{appen:2d}.
The (instantaneous) angular correlation function is
the Fourier transform of the angular power spectrum~$P(\alpha)$:

\begin{eqnarray}\label{eq:cc}
  C(\theta) & = & \int\!\! d^2{\alpha}\, P(\alpha)e^{2\pi i {\boldsymbol
        \alpha}\cdot {\boldsymbol \theta}} \nonumber \\
  & = & \int_0^{\infty}\!\!d\alpha\!\!\int_0^{2\pi}\!\!d\varphi \;\alpha
        P(\alpha) e^{2\pi i \alpha\theta\cos\varphi} \nonumber \\
  & = & 2\pi \int_0^{\infty}\!\!d\alpha \;\alpha P(\alpha)J_0(2\pi \alpha
        \theta).
\end{eqnarray}

To model the temporal variations, we follow the {\em frozen turbulence hypothesis}
(FTH) proposed by \citet{taylor38}, which states that the small-scale turbulence velocity is 
much smaller than the global advection flow. In other words, the atmospheric 
fluctuations are modeled by a frozen pattern of fluctuations with Kolmogorov 
spectrum, blown through the experimental beams by a uniform wind.
This model has been tested and verified by many groups engaged in interferometric observations.
In the {\sc Acbar} observations, a chopping mirror sweeps the beam across the sky. 
In this case, applying FTH requires that the small scale turbulence velocity
be much less than the sum of the chopper and wind angular speeds.

We expect the large scale air flow over 
a flat terrain like the South Pole plateau to be nearly horizontal.
Indeed, \citet{aasto03} report that the vertical wind speed is 
less than several tens of centimeter per second (compared to typical horizontal 
windspeeds on the order of 10 meters per second). In our model, we take
the wind velocity ${\bf v}$ to be completely horizontal.

\section{Observations}\label{sec:observations}
The data used in this study were collected using the Arcminute Cosmology Bolometric Array 
Receiver ({\sc Acbar}, for a detailed description, see \citet{runyan03}): an instrument 
designed to measure the anisotropy of the CMB. 
{\sc Acbar} is an array of 16 bolometric detectors cooled
to $240\,$mK with a 3 stage-$^3$He sorption refrigerator.
The square array is divided into four rows of four detectors each.  The top 
and bottom rows correspond to 274~GHz and 219~GHz (bandwidths 48~and 31~GHz, 
respectively), while the middle two rows are 
sensitive to 150~GHz (bandwidth 31~GHz).  Each detector is offset 
from its nearest neighbors by $\sim 16^{\prime}$.  
{\sc Acbar} is mounted on the Viper telescope, a $2\,$m off-axis Gregorian telescope
at the geographical South Pole. 
The telescope produces beams with FWHM of $4-5^{\prime}$
in all frequency bands. 
Through the use of a chopping flat mirror, the telescope 
sweeps the beams across the sky at frequencies up to several~Hz, with 
peak-to-peak amplitude of $\sim3^\circ$ in the
azimuthal direction. The results presented in this paper are derived from the 
data taken with a chopping frequency of $0.3$ Hz.
{\sc Acbar} has high resolution in spherical harmonic multipole $l$-space 
($\Delta l \sim 100$) over a wide range of angular scales ($150 < l < 3000$), 
allowing precise measurements of the CMB power spectrum (K2004).

The observations used in this work began in late March of 2002 and continued through 
the end of July when the supply of liquid helium at the South Pole station was 
exhausted.  
There are about 200 data files spanning that period, each of which 
constitutes up to six hours of continuous observation.  
One six hour observation consists of approximately 300 ``stares'' each lasting about
one minute. During a stare, the telescope tracks a fixed position on the sky while 
the chopper sweeps out a triangle wave in azimuth.  
Each of these stares falls into one of three categories: lead, main, or trail (LMT).  
Lead and trail stares are offset $\pm 3^\circ$ in RA from the main 
stares and have half the integration time that main stares do.  
Each LMT group of stares is offset in 
declination, so that the final dataset represents a rastered map of the sky.  
\citet{kuo04} produced a CMB angular power spectrum from the
M-(L+T)/2 differenced map. 
However, for the purpose of this atmospheric 
study, LMT stares are not differenced and are treated identically.

\section{Data Processing}\label{sec:data}

The beams of the {\sc Acbar} array are swept in azimuth with a $0.3\,$Hz
triangular oscillation. 
The amplitude of the waveform is $3^\circ$ in azimuth when the telescope is pointed 
at the horizon. 
The timestream from the {\sc Acbar} detectors is binned into vectors, or data strips, 
$s_i^{p}\;,\,i=1,2,...,n$, 
each of which consists of $n=128$ temperature measurements for array element
$p$ covering $3^\circ$ on the sky. 
The data corresponding to sweeps in the right (positive AZ) or 
left (negative AZ) directions are binned separately, resulting in two data vectors
from every complete chopper cycle. 
The data vectors are calibrated in RJ temperature units by referencing them to 
observations of the brightness temperature of Mars and Venus \citep{runyan03}.

\begin{figure}[!bp]
\leavevmode
\plotone{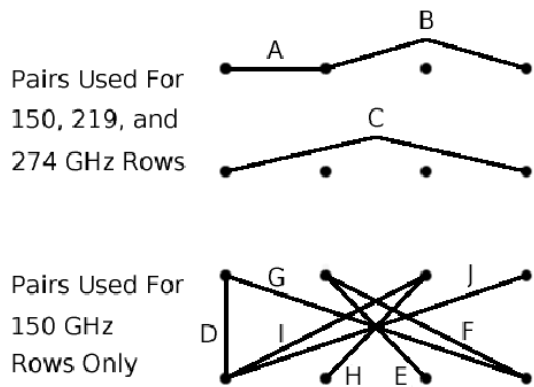}
\caption{Schematic of {\sc Acbar} focal plane demonstrating an example of each array element pair 
correlation that has been used in the analysis.  See Table~\ref{tab:theta} for 
a tabular representation of all possible pairs.  
Correlations are computed for all possible pairs of distinct 
array elements within a frequency passband. The letters A, B, C ... correspond
to the array displacements in Table~\ref{tab:theta}.\label{fig:theta}}
\end{figure}

Correlation matrices are calculated from the outer products of 
data vectors corresponding to different elements in the detector array 
and compared with the model for the atmosphere-induced correlations.  
By correlating data vectors from distinct array elements, we obtain 
cross-correlation matrices that have a significantly lower bias from 
noise power than the autocorrelation.
To maximize the signal-to-noise ratio (SNR), we correlate all 
pairs of array elements within a particular frequency passband.  
The angular offsets between these pairs of array elements 
are listed in Table~\ref{tab:theta}, and a schematic 
diagram of the pairs is presented in Fig.~\ref{fig:theta}.  
Because there are twice as many array elements operating in the 150~GHz band, 
there are many more correlation matrices for that frequency passband.  

The statistical correlation depends only on the relative displacement 
of the pairs of array elements.  
Therefore, we average the correlation matrices of all pairs with the same 
frequency band and displacement 
vector together in a single correlation matrix. 
In order to reduce the volume of data, we assume that the correlation is stationary and 
average the matrices over a period of many sweeps.
The rotation of the telescope with respect to the ground based
wind vector places constraints on the acceptable period of time averaging.
The fundamental length of time over which the correlations are averaged is one quarter of 
a data file or about 1.5 hours (see Section~\ref{sec:chunks} for details).  
This procedure is repeated with the data from the opposite direction 
of chopper motion 
to yield two correlation matrices for each band, detector
displacement vector, and time period. 
After removing modes that are susceptible 
to airmass gradient and chopper synchronous offsets (zeroeth, first, and second 
order polynomial) and subtracting the average of all the stares in the file
from each stare, we compare the measured correlation matrices with those 
calculated from eq.~[\ref{eq:cc}].

\begin{deluxetable}{ccccc}
\tabletypesize{\scriptsize}
\tablewidth{230pt}
\tablehead{
\colhead{Pair} & \colhead{$\Delta\phi$} & \colhead{$\Delta\epsilon$} & 
\colhead{$n_{150}$} & \colhead{$n_{219,274}$}
}
\startdata
A & $-16^{\prime}$ & $0^{\prime}$   & 6 & 3\\
B & $-32^{\prime}$ & $0^{\prime}$   & 4 & 2\\
C & $-48^{\prime}$ & $0^{\prime}$   & 2 & 1\\
D & $0^{\prime}$   & $-16^{\prime}$ & 4 & 0\\
E & $-16^{\prime}$ & $-16^{\prime}$ & 3 & 0\\
F & $-32^{\prime}$ & $-16^{\prime}$ & 2 & 0\\
G & $-48^{\prime}$ & $-16^{\prime}$ & 1 & 0\\
H & $+16^{\prime}$ & $-16^{\prime}$ & 3 & 0\\
I & $+32^{\prime}$ & $-16^{\prime}$ & 2 & 0\\
J & $+48^{\prime}$ & $-16^{\prime}$ & 1 & 0\\
\enddata
\tablecomments{
\label{tab:theta}Angular offset in azimuth $(\Delta \phi)$ and elevation 
$(\Delta \epsilon)$ between array elements used for the correlation analysis.  
The total number of correlation maps 
associated with each array displacement and chopper direction 
is listed for the 150~GHz 
(two rows of detectors) and 219, 274~GHz (one row of detectors each) data.  
The 219 and 274~GHz data contain only displacements in azimuth and have the
same number of correlations maps; $n_{219,274}$ corresponds to either of the
two detector rows.}
\end{deluxetable}

The chopping mirror on the Viper telescope sweeps the optical beams at a 
constant angular speed $\pm\gamma$~(rad~s$^{-1}$).  
At time zero ($t=0$), a detector element measures 
sky temperature $T_1$
in the direction (AZ,\;EL)$=(\phi\;,\epsilon)$. $\tau$ seconds later, 
another detector element (with position offset $\Delta\phi,\;\Delta\epsilon$) 
measures sky temperature $T_2$ in the direction 
of $(\phi',\epsilon')$. The correlation of $T_1$ and $T_2$ is 
$C(\theta)$ given by eq.~[\ref{eq:cc}], where the angular separation $\theta$ is

\begin{eqnarray}\label{eq:theta} 
\theta & = & \sqrt{(\phi'-\phi)^2+(\epsilon'-\epsilon)^2} \\
 & = & \sqrt{(w_\phi\tau\mp \gamma\tau-\Delta\phi)^2+
        (w_\epsilon\tau-\Delta\epsilon)^2}\;,\nonumber 
\end{eqnarray}

and $(w_\phi,w_\epsilon)$ are the components of the wind angular velocity 
in the direction of $(\phi,\epsilon)$. The wind linear velocity is related to
the angular velocity by: 
$$
\frac{v_x}{h}=-\frac{\sin\phi}{\sin\epsilon}w_\phi
+\frac{\cos\phi}{\sin^2\epsilon}w_\epsilon\;,
$$
\begin{equation}\label{eq:wtov}
\frac{v_y}{h}=\frac{\cos\phi}{\sin\epsilon}w_\phi
+\frac{\sin\phi}{\sin^2\epsilon}w_\epsilon\;.
\end{equation}
Note that the angular separation $\theta$ is measured on the ``frozen'' screen of atmospheric 
fluctuations.
When $\Delta\phi=\Delta\epsilon=0$, $C(\theta)$ becomes
the autocorrelation function.
One interesting implication of this scan strategy is that 
the observed atmospheric fluctuation signal depends on the 
direction of the chopper motion (a plus or minus sign in eq.~[\ref{eq:theta}]) 
with respect to the wind direction.  Compared to a drift scan experiment
($\gamma=0$), the fluctuation signal is suppressed when 
the beams and the wind move in the same direction, and enhanced
when they move in the opposite directions.
This effect is easily understood 
in the limit where $\gamma=w_\phi$, $w_\epsilon=0$. 
The optical beams follow the frozen turbulence as it is blown across the sky
and the AC-coupled bolometers will not detect any atmospheric signal.
On the other hand, when the chopper goes in 
the opposite direction to the wind, the instrument will record 
an enhancement in atmospheric signal.

\section{Model Fitting}\label{sec:modfit}

The KT model given by eq.~[\ref{eq:palpha}] is used to characterize the
atmospheric fluctuation angular power spectrum. 
For each observation, we determine a set of parameters that minimizes  
the difference between the theoretical correlation function and the observed data. 
These parameters are defined in eqs.~[\ref{eq:palpha} --  \ref{eq:wtov}] and include 
the amplitude for each observing frequency, $B_{\nu}^2$; a power law exponent, $b$; 
and angular wind velocity in the x and y directions, 
$v_x/h$ and $v_y/h$, respectively. 
As a result of the Earth's rotation, the telescope tracks 
across the sky in azimuth, 
changing by 90$^{\circ}$ during the course of 
a six hour observation.  
Using eq.~[\ref{eq:wtov}], we find the components of the prevailing wind angular
velocity, 
$v_x/h$ and $v_y/h$ as functions of the wind velocity components in the telescope
frame and the time during the observation (which sets $\phi$, the azimuth).

To compare the parameterized model to the data, we first need to generate 
a theoretical correlation matrix, ${\cal C}$, from eq.~[\ref{eq:cc}].  We do this 
by discretizing the temporal lag $\tau$ in eq.~[\ref{eq:theta}] into 
$n$ bins, where $n$ is the length of each data vector being correlated---in 
our case, 128.  If we denote $\tau_i$ as the discretized 
temporal lag associated with the sky temperature measurement in the 
direction of $(\phi\;,\epsilon)$ and $\tau_j$ as the same quantity 
for a distinct detector in the 
direction of $(\phi',\epsilon')$, then the temporal lag between these 
two temperature measurements is $\tau = \tau_i - \tau_j$ and 
we can generate the discrete, 
$n$~x~$n$ correlation matrix ${\cal C}$ from the correlation function as 
follows:

\begin{equation}\label{ijmapping}
{\cal C}^{ij} = C(\theta(\tau, \Delta \phi, \Delta \epsilon)).
\end{equation}
where $\Delta \phi = \phi^{\prime} -\phi$ and $\Delta \epsilon = \epsilon^{\prime} - \epsilon$.
We perform an equivalent mode removal on ${\cal C}^{ij}$ 
as was performed on the observed data, so that the theoretical and observed 
correlations can be directly compared.
We calculate a $\chi^2$ for the fit of the theoretical correlation matrix to the observed data
for each frequency passband and unique correlation matrix:  

\begin{equation}\label{eq:chisq}
\chi^2_{k,\nu} = \frac{\sum_{i,j} ({\cal C}_{k,\nu,model}-{\cal C}_{k,\nu,data})^2}{\sigma_{k,\nu}^2},
\end{equation}

\noindent where $\sigma_{k,\nu}$ represents the intrinsic detector noise associated with the 
$k^{th}$ correlation map for observation frequency $\nu$. 
We compute the correlation matrices corresponding to the noise in a given 
6 hour observation 
(denoted by ${\cal C}_{k,\nu,noise}$) as described in Section~\ref{sec:data} with the 
difference that the data vectors for the detector pairs are drawn from the three 
possible pairs of 1.5 hour subfiles separated by at least 1.5 hours in time.  
We expect the atmosphere to be uncorrelated between these two measurements and 
the correlation should provide an accurate estimate of the noise.
Finally, $\sigma_{k,\nu}$ is obtained from the variance of the $k^{th}$ correlation 
matrix of frequency $\nu$:

\begin{equation}\label{eq:chisqnoise}
\sigma^2_{k,\nu} = \sum_{i,j}({\cal C}_{k,\nu,noise})^2
\end{equation}

The overall $\chi^2$ used to determine the model parameters is given by 

\begin{equation}\label{eq:chisqovr}
\chi^2 =  \frac{\sum_{k,\nu} \chi^2_{k,\nu} w_{k,\nu}} { \sum_{k,\nu} w_{k,\nu}},
\end{equation}

\noindent
 where $w_{k,\nu}$ represents the weight of the $k^{th}$ correlation matrix 
found in column 4 ($\nu=150 \;$GHz) or 5 ($\nu=219 \;$or $\;274\;$GHz) of Table~\ref{tab:theta}.  
For example, $k=0, \; \nu=274 \;$GHz corresponds to left- or right-going correlation matrices with 
angular offsets of $\Delta\phi=-16^{\prime}$ and $\Delta \epsilon =0^{\prime}$.  From Table~\ref{tab:theta}, 
we see there are three pairs with this angular offset, meaning $w_{0,274} = 3$.  
$k=1,\nu=150 \;$GHz corresponds to left- or right-going correlation matrices with 
angular offsets of $\Delta\phi=-32^{\prime}$ and $\Delta \epsilon =0^{\prime}$, 
so $w_{1,150} = 4$.

\begin{figure*}[!htb]
\begin{center}
\leavevmode
\epsscale{0.9}
\plotone{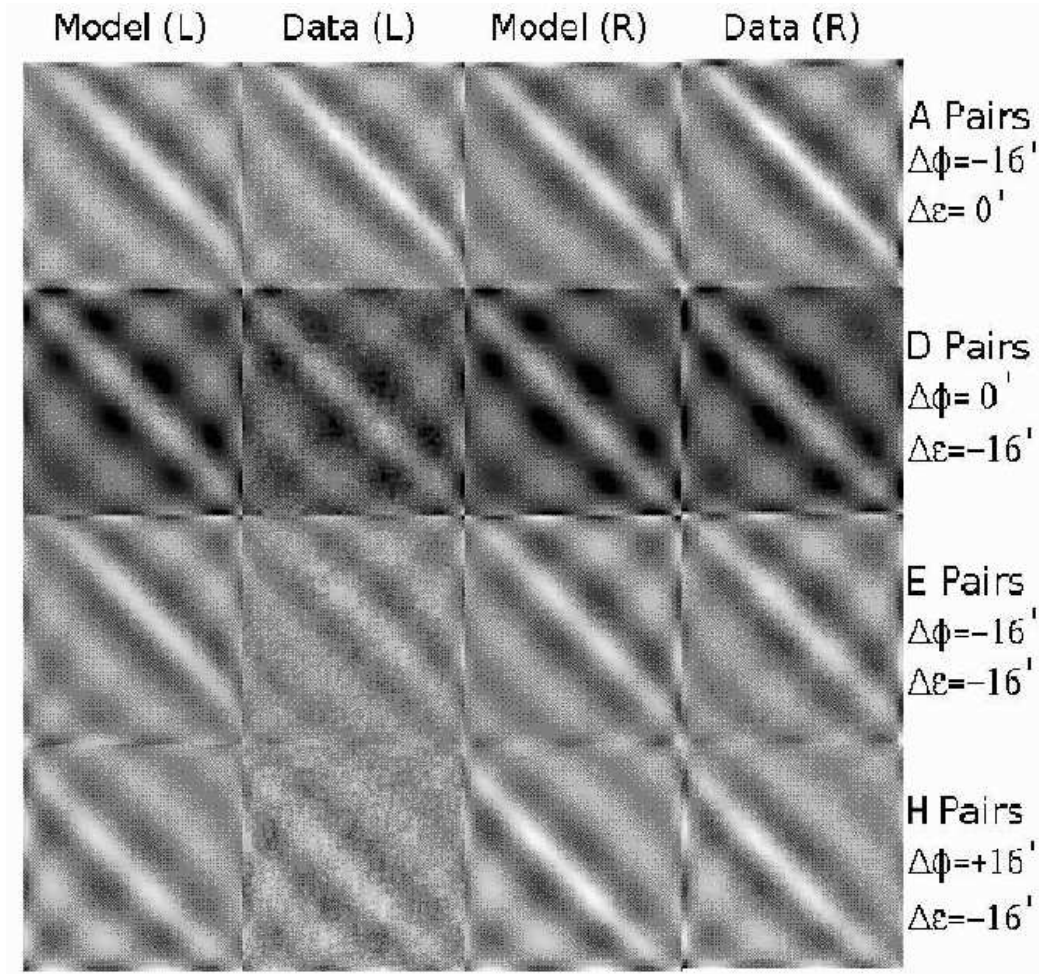}
\end{center}
\caption{Correlation matrices produced by the model (1st and 3rd columns) as compared to 
the observed correlation matrices (2nd and 4th columns) for a 1.5 hour file 
of $150\,$GHz data from May 30, 2002. The two columns on the left 
correspond to a left-going motion of the chopper 
while the two columns on the right correspond to a right going sweep.
The rows are a representative sampling of available angular 
displacements between $150\,$GHz array elements.
\label{fig:samplecorrb}}
\end{figure*}

The lack of detector differences in the $\epsilon$ direction for the 219 and $274\,$GHz data 
results in a poor determination of this component of the wind velocity 
from this data alone. 
However, the 150~GHz data feature correlation matrices with angular offsets in both 
$\phi$~and~$\epsilon$ directions, and can be used to obtain reliable wind velocity parameters. 
We simultaneously fit the data for all frequencies with the 
wind velocity to be constrained to be the same for all detector pairs.
The $150\,$GHz data provides constraints on the $\epsilon$ component of 
the wind velocity. 
We use the IDL AMOEBA routine to perform the minimization and determine 
the best fit parameters for the KT model. 

\subsection{Correlation Matrix Fits}\label{sec:corrmat}

\begin{figure*}[!htb]
\begin{center}
\leavevmode
\epsscale{0.8}
\plotone{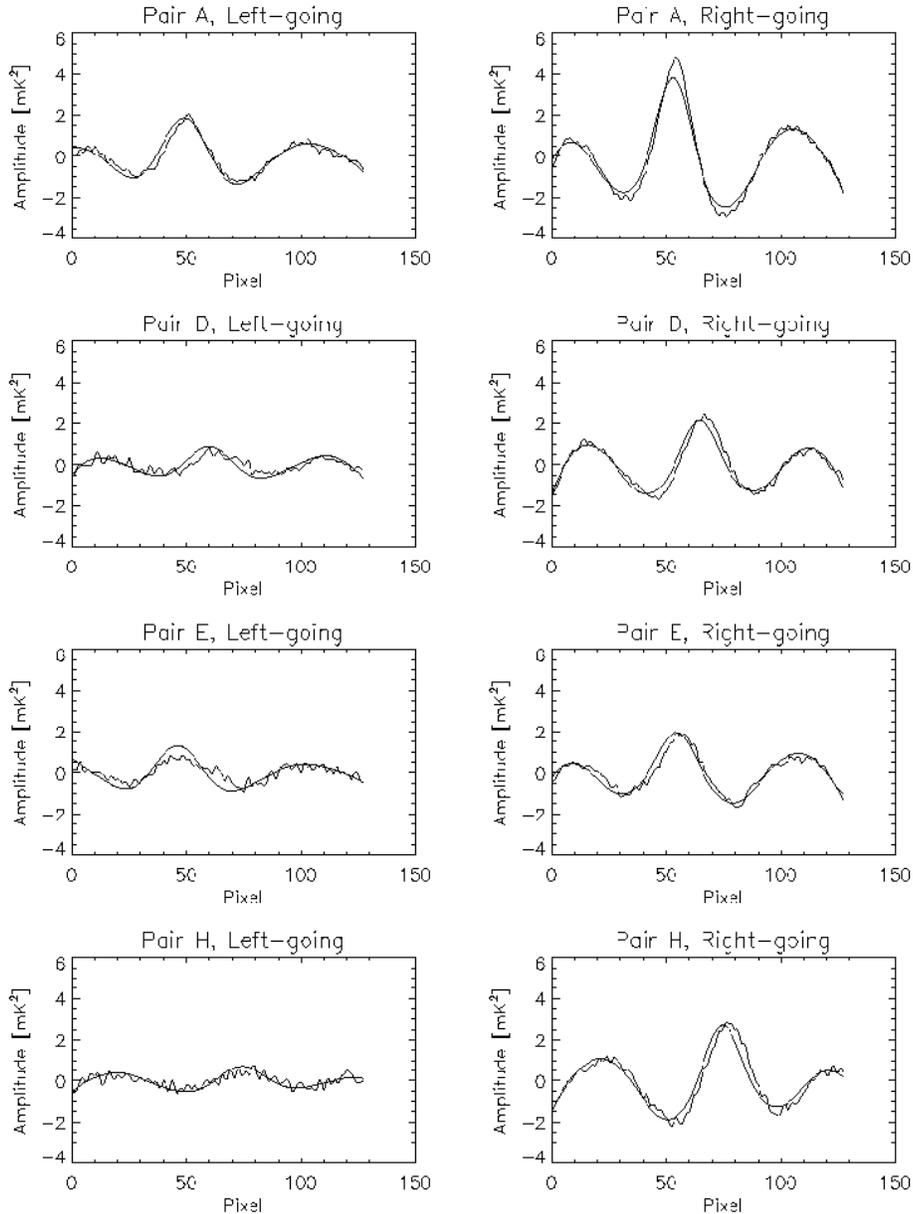}
\end{center}
\caption{Cross-Sectional plots of the observed correlation matrices shown in 
Fig.~\ref{fig:samplecorrb} with the model fit overplotted.  The cross-section is 
taken in the horizontal direction through the center of the correlation matrix, 
and serves as a demonstration of the quality of the fit.  
Due to the high wind velocity, the sky noise is suppressed for the 
chopper moving in the same direction as the wind (left column) 
or enhanced for motion in the opposite direction 
(right column).\label{fig:samplecorra}}
\end{figure*}

We begin by fitting the model parameters to the data divided in 1.5 hour
segments. 
We find that the atmospheric fluctuations observed by {\sc Acbar} are
well described by eqs.~[\ref{eq:cc}, \ref{eq:theta}] for about 93\%
of the data.
A typical example of 
a good fit by the model is shown in Figs.~\ref{fig:samplecorrb}~and
~\ref{fig:samplecorra}.  These figures show the 
theoretical and observed correlation matrices for a 1.5 hour data file obtained on 
2002 May 30.  
The two columns on the left correspond to left-going sweeps of the telescope, 
while the two columns on the right represent the opposite direction.  
Fig.~\ref{fig:samplecorra} shows a cross-section through the center 
of the correlation matrices presented in Fig.~\ref{fig:samplecorrb}.  
This file has a particularly high azimuthal component of wind velocity and the cross-section 
plots demonstrate the difference in amplitude between the left-going and 
right-going sweeps.  As described in Section~\ref{sec:data}, 
the sky noise is suppressed when
the beam follows the wind (as seen in the left-going sweeps for this day), 
and enhanced when they move in the opposite directions.

\begin{figure}[!htb]
\begin{center}
\leavevmode
\epsscale{1.0}
\plotone{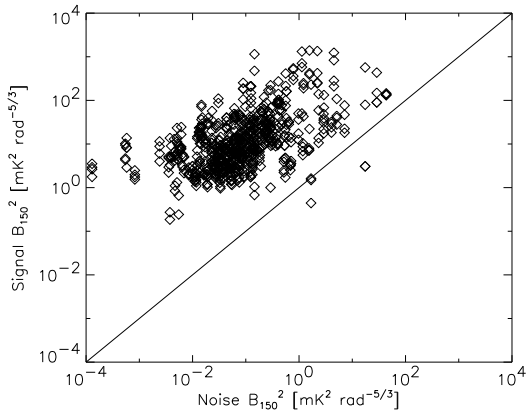}
\end{center}
\caption{
A graphical representation of the SNR in the determination of the fluctuation
power amplitude.
On the y-axis, we plot the mean fluctuation amplitude found from
the individual 1.5-hour segments of a 6 hour data file.
The x-axis is the amplitude computed from
the correlation between detector timestreams separated by at least 1.5 hours
within a given 6-hour observation. 
The diagonal line represents equal fluctuation power amplitude 
for the signal and noise correlations.
The fluctuation amplitude is recovered with high SNR in nearly all files.
In fact, the median SNR is 140.
\label{fig:sig_to_noise}}
\end{figure}

One of the unique features of this work is that it appears possible---provided 
the KT model reasonably describes the data---to measure the 
atmospheric fluctuation power under nearly all atmospheric conditions, 
even during the best weather.  
As mentioned in Section~\ref{sec:modfit}, the intrinsic detector noise sets a 
fundamental limit to the smallest atmospheric fluctuation power that 
can be measured. 
We characterize the noise in the fluctuation amplitude by fitting to the
random correlations between detectors in time-separated sweeps.
In Fig.~\ref{fig:sig_to_noise}, we plot the noise power amplitude 
vs. the power amplitudes found for the correlations 
within the sweeps. 
The line represents a SNR of approximately one.
Nearly all of the data lie above this line, indicating that the amplitude of the 
fluctuations is not limited by an instrumental noise floor and that the amplitude
of the fluctuation power is typically measured with high accuracy 
(median SNR is 140) under nearly all observing conditions. 

We identify the observations that do not provide adequate fits to the KT model 
from the $\chi^2$ defined by eq.~[\ref{eq:chisqovr}].
We find that 7\% of the total observations fail to measure the 
KT model amplitude with $\chi^2 < 20*median(\chi^2)$.
In general, the poor fits are not due to a lack of fluctuation power, but rather 
the presence of fluctuations that cannot be described by our assumed model.
We have broken these files into smaller time periods and find no 
improvement in the fits, 
indicating that it is not a problem with the fit parameters changing in time. 
Two possible explanations are that either the observed spatial fluctuations do not obey a 
Kolmogorov power law at this time, or the atmospheric structure is not frozen on the 
timescale of a sweep.
Another more likely explanation is that a significant fraction of the observed power is 
coming from a layer in the 
near field of the telescope, which we estimate to be approximately $500\,$m.
In this case, the assumptions made in the derivation of the theoretical  correlation 
function are not valid.
The data with poor fits are typically associated with very large fluctuation amplitudes,
and are therefore likely to be due to a severe storm.  

\subsection{Timescale of Disturbances}\label{sec:chunks}

By averaging the correlation matrices over longer periods of time, it is 
possible to measure the model parameters (particularly the wind velocity) with higher 
accuracy.
However, it is important that the atmospheric conditions remain 
constant over the period for which the correlations are averaged.
The CMB observations we are using for this analysis cover 6 hours;
this sets a practical upper limit to the averaging.
In this section, we show that the atmospheric conditions are typically
quite stable over the course of a six hour observation.

\begin{figure}[!bp]
\begin{center}
\leavevmode
\epsscale{1.0}
\plotone{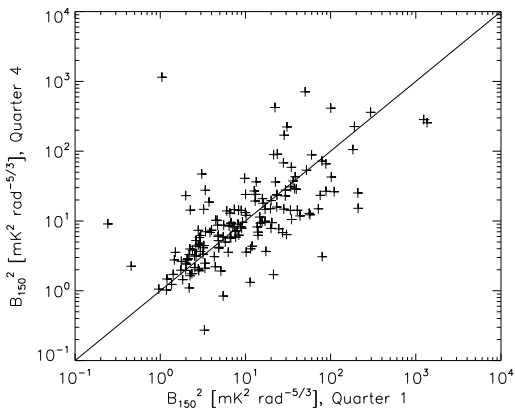}
\end{center}
\caption{Correlation of brightness fluctuation power amplitude over the span of 
a six hour observation.  We show correlation between $B_{150}^2$ 
measured from first and fourth quarters of a six hour data file.
The strong correlation implies that that brightness
fluctuation power typically remains constant over a time period of at least six hours. 
\label{fig:amp_q}}
\end{figure}

Each of the six hour data files is subdivided into four 1.5 hour subfiles. 
Model parameters are determined for each of the subfiles from fits to the 
$150\,$GHz data.
To reduce computing time, we fit only to the $150\,$GHz data for the tests described 
here and in Section~\ref{sec:spindex}. 
In Fig.~\ref{fig:amp_q}, we show the correlation between the power amplitudes
determined for the 1st and 4th quarters of each file. 
The strong correlation indicates that the atmospheric fluctuation
power is relatively consistent in time and we are justified in fitting a single
fluctuation power to a six hour data set.
Due to the rotation of the telescope with respect to the prevailing wind 
direction as it tracks a source, we cannot simply average the correlation
matrices over the entire 6 hour observation. 
Instead, we project a constant ground-based wind velocity into  
the appropriate telescope frame for each of the quarters of the file.
The minimization for the model parameters then proceeds as for a single file;
however, we use the combined $\chi^2$ for all four files as a metric of goodness of fit.   

\newpage

\subsection{Power Law Exponent}\label{sec:spindex}
\begin{figure}[!bp]
\begin{center}
\leavevmode
\epsscale{1.00}
\plotone{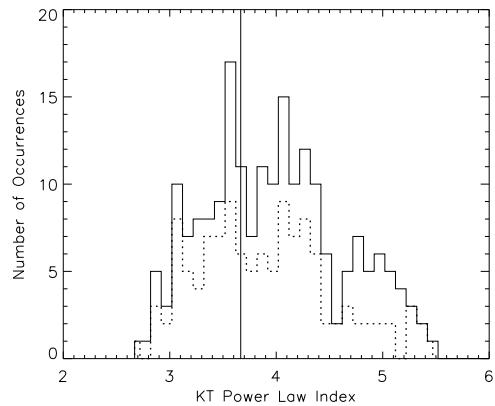}
\end{center}
\caption{Distribution of the measured KT 
power law  exponent values for the Austral winter of 2002.  Observations that do not fit the 
KT power law (according to the $\chi^2$ cut described in Section~\ref{sec:corrmat}) 
are not included in this distribution.  
The solid line shows the distribution of exponents for all good fits to the KT
model; the mean of this distribution is 4.1$\pm$0.8. 
The dotted lines shows the distribution of exponents for all observations with 
below-mean $\chi^2$ for the fit to the KT model; 
the mean of this distribution is 3.9$\pm$0.6. 
In both cases, the results agree with the theoretical KT power law for small angular 
scales, which is shown as a vertical line.  
For the rest of the analysis, we fix the exponent at the expected value of 
$b=11/3$ to remove the degeneracy between the amplitude and exponent of the 
power law.\label{fig:exponent}}
\end{figure}

We can determine the exponent of the KT power law  
for those observations in which we measure the fluctuations
with high SNR. 
In Fig.~\ref{fig:exponent}, we show the distribution of measured 
values of the power law exponent over the course of the winter where  
the mean exponent for that period of time is $b=4.1\pm0.8$.
If we restrict the observations to those with above-median SNR, we find that the 
distribution of exponents is described by $b=3.9\pm0.6$.
In both cases, our results are consistent with those expected for the KT
model in the 3D regime. 
For the rest of our analysis, we have set the exponent of the KT power
law equal to 11/3.
This assumption is physically motivated and increases the stability of the 
model-fitting by removing one of the free parameters.  

\section{Results}\label{sec:results}

In this Section, we describe the basic results of the fits of the KT
model to the observed data. 
As described in Section~\ref{sec:chunks}, the stability of the atmospheric 
conditions allows us to determine the model parameters for an entire
six hour observation. 
Although the fluctuation amplitude is determined with high SNR 
in each of the file quarters, fitting to all four simultaneously significantly 
improves the determination of the KT exponent and wind velocity components. 
Unless otherwise noted, the following results are found assuming that
the model parameters are constant over a six hour data file.
The amplitudes for the fluctuation power are free to vary in each of the bands, 
while the wind angular velocity is constrained to be constant across all bands.
For the 219 and 274 GHz rows, this greatly improves the stability of the results. 

\subsection{Amplitude Measurements}\label{sec:ampmeas}

\begin{figure}[!bp]
\begin{center}
\leavevmode
\epsscale{1.00}
\plotone{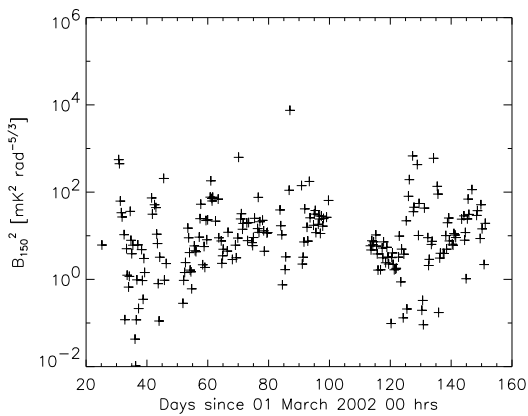}
\end{center}
\caption{The brightness fluctuation power, in mK$^2$~rad$^{-5/3}$, for 4 months of 
150~GHz {\sc Acbar} observations at the South Pole.  Each data point represents 
about six hours of continuous observations normalized to observations at the 
zenith angle.  
Approximately 7\% of 
the data exhibit anomalous turbulence which are not well fit by a KT
model, and are not shown here.  We assume that these data always have amplitudes 
larger than data which can be modeled with a KT power law and adjust our 
amplitude distributions accordingly.
Mechanical problems with the instrumentation limited the acquisition of 
data for days 100-112.
\label{fig:amptime}}
\end{figure}

We have used fits to the KT model to measure the 
brightness fluctuation power amplitude of the atmosphere over 
a four month period from late March through July 2002. 
Although the measurements were made at a range of elevations, the results have been normalized 
in terms of observations at the zenith.
The results of this analysis for 150~GHz data are 
shown in Fig.~\ref{fig:amptime}.
The atmospheric brightness fluctuation power is relatively 
stable, with roughly 60\% of the data falling between 
$B_{150}^2 = 5 - 40\,{\rm mK}^2\,{\rm rad}^{-5/3}$.  
This is in contrast to the findings of LH2000, who reported a bimodal 
distribution of brightness fluctuation power for the Austral 
summer of 1996.  Solar heating and associated weather during the summer is 
one possible cause of this difference.

\begin{figure}[!tp]
\begin{center}
\leavevmode
\epsscale{1.0}
\plotone{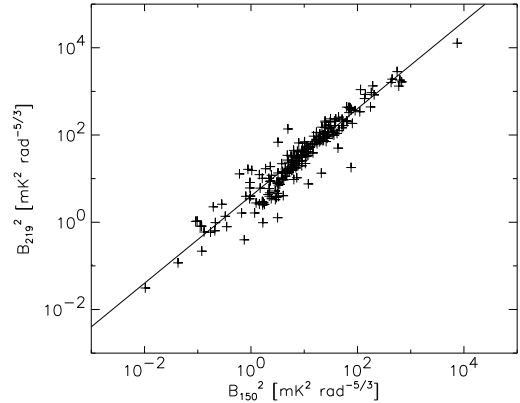}
\end{center}
\caption{The fluctuation power amplitudes determined at $150\,$GHz vs. $219\,$GHz.
The high degree of correlation is a result of the large SNR 
in the determination of the amplitudes and the robust fits to
the model. A similar plot for the $150\,$GHz vs. $274\,$GHz results
is qualitatively identical. \label{fig:150_219}}
\end{figure}

\begin{figure}[!bp]
\begin{center}
\leavevmode
\epsscale{1.0}
\plotone{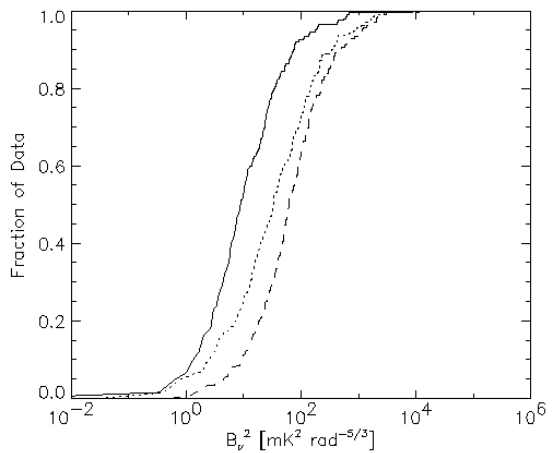}
\end{center}
\caption{The cumulative distribution functions of brightness 
fluctuation power amplitude (in Rayleigh-Jeans temperature units) 
for the entire winter season of {\sc Acbar} observations. 
The solid, dotted, and dashed lines 
represent 150~GHz, 219~GHz, and 274~GHz data, respectively.  
The 7\% of the data that is poorly fit by a KT model is not included in 
this plot.
The upper limits to amplitude quartiles given in Table~\ref{tab:quartiles} 
have been calculated from these distributions.
\label{fig:cdfamp3}}
\end{figure}

We have performed a similar analysis for the 219 and $274\,$GHz data.  
The emissivity of water vapor increases as a function of frequency, resulting in 
higher amplitude fluctuations at higher frequencies. 
In Fig.~\ref{fig:150_219}, we compare the amplitudes
found from the $219\,$GHz observations with those found at $150\,$GHz. 
The degree of correlation is a consequence of the high SNR of all 
the measurements and is similar to that found between the 150 and $274\,$GHz data.
The brightness fluctuation power amplitude at 274 (219) GHz is 
approximately 7.4 (3.7) times greater than at $150\,$GHz.
In Fig.~\ref{fig:cdfamp3}, we show a Cumulative Distribution Function (CDF) of 
the brightness fluctuation power for each of the three frequency bands.  
From these distributions, we extract upper limits to the quartiles in 
brightness fluctuation power amplitude and list them in Table~\ref{tab:quartiles}.  
We compensate for the 7\% of the data that do not fit the KT model 
by assuming that all the poor KT fits have amplitudes larger than the largest
good KT fit. 
The 27th, 54th, and 81st percentiles then serve as upper limits to the quartiles 
of the full distribution.
 
The median fluctuation power amplitudes for each of the bands are listed in 
Table~\ref{tab:multifreq}.
For comparison, we also list the median fluctuation power measured by the 
Python experiment from the paper of LH2000.  
Due to the increase in atmospheric brightness with frequency and the higher 
sensitivity of the instrument, the {\sc Acbar} measurements are significantly 
more sensitive to water vapor than those made with Python.
This sensitivity is helpful in characterizing the much more quiet
winter atmosphere.

\begin{figure}[!bp]
\begin{center}
\leavevmode
\epsscale{1.0}
\plotone{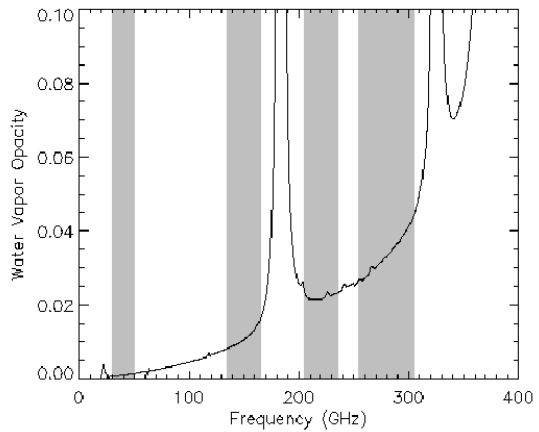}
\end{center}
\caption{Water vapor emission spectrum obtained from the difference 
between a  ``wet'' spectrum with 0.25~mm input water vapor and a 
``dry'' spectrum with 0.0~mm water vapor.  Approximate frequency passbands of 
the Python (40~GHz) and {\sc Acbar} (150, 219, 274~GHz) experiments
are shown with grey rectangular bars. \label{fig:wvop2}}
\end{figure}

We have used a model for the spectrum of atmospheric water vapor to compare
the results from the four frequency bands covered by {\sc Acbar} and Python. 
The Atmospheric Transmission (AT)\footnote{Airhead Software, Boulder CO} 
code is used to produce spectra corresponding to winter conditions at 
the South Pole with 0.25~mm water vapor as well as 0.0~mm water vapor.  
The difference of these two spectra gives an estimate of the emission due to 
water vapor as a function of frequency.
The emission spectrum of atmospheric water vapor is shown in Fig.~\ref{fig:wvop2}, 
with the four passbands shown in gray.  
We integrate over the passbands to obtain the effective opacity $\tau^{\prime}_{\rm band}$ 
due to $0.25\,$mm water vapor for each of the observing frequencies; 
40, 150, 219, and 274~GHz.
We then scale the fluctuation power amplitude for each of the bands by 
the square of the ratio of the water vapor opacities to normalize 
the results to the equivalent power that would have been seen in simultaneous 
observations at $150\,$GHz.
$$ {\rm Equivalent} \;B^2_{150} =  B^2_{\nu} \left(\frac{\tau^{\prime}_{150}}{\tau^{\prime}_{\nu}}\right)^2$$  
Given a perfect understanding of the {\sc Acbar} bands,
an accurate water vapor spectrum, and the validity of the assumption
that the atmosphere is optically thin, we expect the scaled 
fluctuation amplitudes for the {\sc Acbar} measurements to be identical.
The equivalent brightness fluctuation power amplitudes for the Python observations
made during the Austral summer are more than 10 times higher than those for the 
{\sc Acbar} measurements made during the Austral winter.
We attribute this difference largely to the lower water vapor content of the atmosphere 
during the winter, although the stability of the atmosphere may also be 
improved in the winter.

The measurements of the fluctuation amplitude made with the Python experiment and
reported in LH2000 are given in terms of $Ah^{8/3}$, while the {\sc Acbar} 
results are given in terms of $B_{\nu}^2 = Ah^{5/3}$.  
In order to compare the results of the experiments,
it is necessary to divide the numbers from LH2000 by the altitude of the layer 
containing the fluctuations.
This additional factor of $h$ was introduced by the authors of LH2000 so that 
the data would be easier 
to insert into a set of approximate analytic expressions predicting the level of residual 
atmospheric noise in different classes of observations. However, these expressions 
are only valid in the limit of long averaging times, which are not appropriate for
many observing strategies. $B_{\nu}^2$ is what is actually measured 
by the experiments, and is therefore more fundamental and useful than $Ah^{8/3}$. In 
Section~\ref{sec:windmeasure} we show that the scale height of the fluctuations 
is consistent with the distribution of precipitable water vapor.  
Should one desire to compute $Ah^{8/3}$ from the {\sc Acbar} data, multiply $B_{\nu}^2$
by the water vapor pressure weighted altitude, $h=1300\,$m.     

\begin{deluxetable}{cccc}[!htb]
\tabletypesize{\scriptsize}
\tablewidth{230pt}
\tablehead{
\colhead{Percentile} & \colhead{27th} & \colhead{54th} & \colhead{81th}}
\startdata
150~GHz & 3.7 & 10. & 37.
\\
219~GHz & 11. & 38. & 160
\\
274~GHz & 28. & 74. & 230
\\
\enddata
\tablecomments{
\label{tab:quartiles}Upper limits to quartiles of fluctuation amplitude 
($B_{\nu}^2$) for 150, 219, and 
274 GHz data in Rayleigh-Jeans temperature units of mK$^2$ rad$^{-5/3}$.  
These values are calculated from the cumulative distribution 
functions shown in Fig.~\ref{fig:cdfamp3}.  We report the 27th, 54th, and 81st 
percentiles, 
because the distribution from which they are calculated does not include the 7\% 
of non-KT observations.  We assume that each of the non-KT observations 
has an amplitude larger than the largest KT observation, the values in this table 
represent upper limits to the 25th, 50th, and 75th percentiles of the full distribution.
}
\end{deluxetable}

\begin{deluxetable*}{lccccc}[!htb]
\tabletypesize{\scriptsize}
\tablewidth{350pt}
\tablehead{
\colhead{Frequency} & \colhead{40~GHz} & \colhead{150~GHz} & 
 \colhead{219~GHz} & \colhead{274~GHz}
}
\startdata
Median $B_{\nu}^2$ [mK$^2$~rad$^{-5/3}$] (CMB units) & 0.90 & 31. & 350 & 2.0x10$^3$
\\
Median $B_{\nu}^2$ [mK$^2$~rad$^{-5/3}$] (RJ units) & 0.84 & 10. & 38. & 74.
\\
Equivalent $B_{\nu}^2$ for $150\,$GHz & 11. & 1.0 & 0.92 & 0.86
\\
\enddata
\tablecomments{
Comparison of median brightness fluctuation power for measurements
with the {\sc Acbar} and Python (see LH2000) experiments.
The results are listed in both Rayleigh-Jeans and CMB temperature units.
Clearly, RJ units are more natural for comparing between different bands,
however, we include the CMB normalization in order to facilitate the prediction  
of noise in future experiments.
We use a water vapor spectrum to normalize the 
atmospheric fluctuation power for the Python $40\,$GHz and {\sc Acbar} 219, $274\,$GHz 
observations to the equivalent power for the $150\,$GHz {\sc Acbar} band.
The data for the three {\sc Acbar} bands were taken simultaneously and, therefore,
the equivalent $150\,$GHz amplitudes should be roughly identical.
The Python numbers are computed from Table 1 in LH2000 assuming the average altitude 
of the fluctuations, $h_{av} = 1000$m. 
The order of magnitude higher equivalent  
fluctuation power for the extrapolated $40\,$GHz Python results  
is a result of the lower precipitable water vapor and increased stability of the 
atmosphere during the winter at the South Pole.
}\label{tab:multifreq}
\end{deluxetable*}

\begin{figure}[!bp]
\begin{center}
\leavevmode
\epsscale{1.0}
\plotone{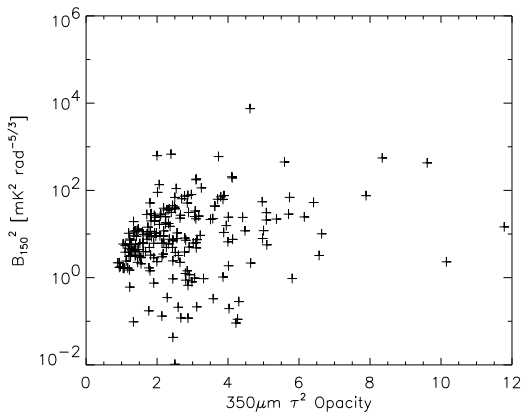}
\end{center}
\caption{Fluctuation power amplitude versus $350\,\mu$m opacity-squared.
The weak correlation between $B_{150}^2$ and $\tau^2$ implies that $\tau$ 
alone is not a reliable indicator of atmospheric fluctuations. \label{fig:tauamp}}
\end{figure}

Measurements of the sub-mm atmospheric opacity are an important monitor of 
observing conditions.  We have investigated the correlation of the measured 
fluctuation amplitude with both the sub-mm opacity ($\tau$) and the 
variance of the opacity ($\sigma_\tau$).  
The sub-mm tipper (sky brightness monitor) on the AST/RO experiment 
\citep{radford02a} at the South Pole measures the $350\,\mu$m zenith opacity 
and atmospheric temperature every 15 minutes.
The $350\,\mu$m atmospheric opacity is much larger than $150\,$GHz opacity, 
but they are both dominated by water vapor and, therefore, scale nearly linearly.
We bin the opacity values from the tipper to match the 
length of the observation (typically about six hours); the fractional variance 
in opacity over that time-span ($\sigma_\tau / \tau$) is usually less than 10\%. 
Fig.~\ref{fig:tauamp} shows the fluctuation power amplitude versus the 
square of the opacity, $\tau^2$. 
The correlation between the fluctuation power amplitude and the variance of the
opacityis similarly weak. 
These results imply that $\tau$ and $\sigma_\tau$ 
are not, by themselves, an adequate judge of the stability 
of the atmosphere on timescales relevant to the {\sc Acbar} measurements.  

\subsection{Wind Velocity and Scale Height}\label{sec:windmeasure}

In general, the angular speed of the chopper is higher than the effective angular
speed of the fluctuations produced by wind. 
Therefore, the effect of the wind velocity on the observed correlations is typically 
sub-dominant to the effect of the chopper motion. 
In practice, the amplitude of the fluctuations can usually be recovered with high
SNR even when very poor constraints can be placed on the wind velocity. 
However, in cases where the fluctuations are measured with high SNR,
the wind angular velocity can be accurately recovered.  
This wind angular velocity can be combined with radiosonde measurements of 
wind linear velocity to estimate the scale height from which the atmospheric 
fluctuation power originates.  

Data files with a $\chi^2$ (as described by eq.~[\ref{eq:chisqovr}]) below the median 
are selected as being likely to result in reliable wind velocities.
The wind angular velocities for each 1.5-hour file are found in a coordinate system 
that is fixed with respect to the direction the telescope is pointing.
However, over the course of a long data file, the telescope tracks a source through 
a considerable range of angle and this motion must be taken into account. 
The transformation described by eq.~[\ref{eq:wtov}] ties the local wind 
velocities for each 1.5-hour data set into a single wind velocity with 
respect to the ground constant over the entire data file. 
We are justified in doing this because both the measurements of the wind angular
velocities with {\sc Acbar} and the wind linear velocities with radiosondes show that
the wind is typically quite constant over a 6-hour data file.

The {\sc Acbar} measurements of wind angular velocity can be combined with 
radiosonde linear velocity and water 
vapor pressure as a function of elevation to determine the mean elevation ($h$) 
from which the fluctuations originate: 

\begin{equation}\label{eq:scheight}
h = \frac{v_{rs}}{\omega_{Acbar}}
\end{equation}

\noindent where $v_{rs}$ is defined as 

\begin{equation}\label{eq:vrad}
v_{rs} = \frac{\int^{\infty}_0 v(z)P_{wv}(z)dz}{\int^{\infty}_0 P_{wv}(z)dz}\,,
\end{equation}

\begin{figure}[!bp]
\begin{center}
\leavevmode
\epsscale{1.0}
\plotone{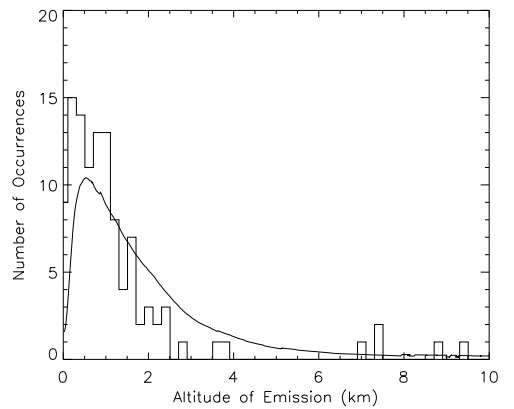}
\end{center}
\caption{Distribution of the inferred altitudes for the atmospheric fluctuations. 
These results are computed from the ratio of the {\sc Acbar} (below-median $\chi^2$ 
$150\,$GHz data) 
wind angular velocities to the radiosonde derived water vapor pressure 
weighted wind linear velocities.  The results are consistent with the 
average water vapor pressure profile calculated from the data shown in 
Fig.~\ref{p_wv} and reproduced here as a smooth line.
The mean inferred scale height of the emission is $\sim 1.3\,$km.\label{fig:windalt}}
\end{figure}

\noindent $v(z)$ is the wind linear speed from the radiosonde data, 
$P_{wv}(z)$ the water vapor pressure (calculated in Section~\ref{sec:genprop}), and 
$z$ the altitude above ice level.
Fig.~\ref{fig:windalt} is a histogram of the altitude of the inferred layer giving rise to the
measured fluctuations.
The results are in reasonable qualitative agreement with the 
average water vapor pressure weighted altitude calculated from radiosonde data
shown in Fig.~\ref{p_wv}. 
This provides further confidence that we understand the source of the fluctuations
and have accurately characterized them.

\section{Conclusions}\label{sec:conclusions}
We have produced a detailed characterization of the millimeter wavelength 
atmospheric fluctuations at the South Pole during the Austral winter of 2002. 
By correlating the signals between pairs of elements in the {\sc Acbar} 
bolometer array it is possible to reduce the effect of uncorrelated detector noise
such that the atmospheric fluctuations can be measured with high SNR 
even for the lowest fluctuation amplitudes.  
We found that a KT model is an accurate description of the 
atmospheric fluctuation power $\sim 93\%$ of the time. 
The median exponent of the power law is found to be consistent with the 
theoretically expected value $b=11/3$.
For the bulk of the analysis, we fix the exponent to 11/3 in order to improve
the SNR in the determination of the other model parameters.
We found upper limits to the the quartiles of the $150\,$GHz atmospheric fluctuation 
power for the four winter months of observation to be 
[3.7, 10., 37]~mK$^2$~rad$^{-5/3}$.  
The scaling between the measured fluctuation power amplitudes for each of the three {\sc Acbar} frequencies
is consistent with that expected for emission from water vapor.
We used this fact to compare our results with those of LH2000 made at $40\,$GHz with
the Python experiment in the Austral summer of 1996.
Assuming that the scale heights of the fluctuations in the summer and winter are 
equivalent, the median fluctuation power found with {\sc Acbar} observations in the 
Austral winter corresponds to 
a fluctuation power amplitude $\sim 10$ times smaller than that found with Python during the 
Austral summer.  
From the analysis in LH2000, this implies that the atmospheric fluctuation power at the 
South Pole in the Austral winter is at least $\sim 30$ times smaller than that at the 
ALMA site in the Chilean Andes. 
In addition, we found that the fluctuation power amplitude is only weakly correlated with 
atmospheric opacity or its variance on 15 minute timescales.
Therefore, for experiments sensitive to atmospheric fluctuations, periodic measurements of 
optical depth do not appear to be an adequate characterization of an astronomical site.
Finally, we determined the wind angular velocity for each of the observations with 
below median $\chi^2$ and compared our value to the wind linear velocities obtained 
by radiosonde 
measurements to determine the scale height of the observed atmospheric fluctuations.  
We find that the inferred scale heights for each of the data files roughly fall within 
the $300\,$m--$2000\,$m range suggested by the radiosonde water vapor pressure profiles.

We believe this work provides a fairly complete picture of the fluctuation power in the atmosphere
above the South Pole. Already these results are being used to compare potential observation
strategies for the 10-meter South Pole Telescope, which will be deployed in 2007. 
We anticipate that
similar studies to that presented here, will be performed with data from other
array receivers in order to 
compare quantitatively the atmospheric stability of observing sites.

The authors gratefully acknowledge the contributions of the entire {\sc Acbar}/Viper team in 
the construction, maintenance, operation, and calibration of the experiment. 
We would also like to thank
Nils Halverson and Tom Crawford for their careful reading of an early draft of this 
work and their helpful suggestions for it's improvement. 
This work has been supported by the NSF office of polar programs grant OPP-0091840.

\appendix

\section{Projected Power Spectrum}\label{appen:2d}

\begin{figure}[!htb]
\begin{center}
\leavevmode
\epsscale{0.5}
\plotone{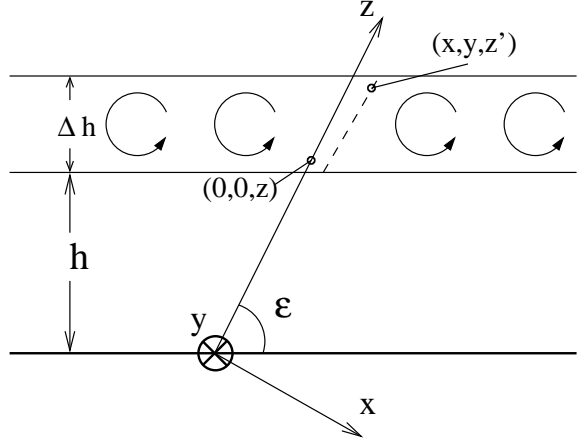}
\end{center}
\caption{The coordinate system for the angular power spectrum 
derivation given in the appendix.\label{fig:coord}}
\end{figure}

Here we derive the instantaneous angular power spectrum, given the 
three dimensional temperature-density fluctuation spectrum.
The thermodynamic temperature of the atmosphere is on the order of
$200\,$K, hence the Rayleigh-Jeans law is a good approximation at 
mm wavelengths. In the optically thin limit and under the condition 
of local thermodynamic equilibrium, the specific radiation 
is proportional to the line of sight integration of the light emitting 
elements. Therefore, we can define the effective temperature 
$T_E({\bf r})$, such that the brightness temperature $T_B$ along 
${\bf \hat r}\equiv {\bf r}/r$ is given by
$$
T_B=\int T_E \kappa\,dr\;,
$$
where $\kappa$ is the opacity/thickness of the turbulent layer for millimeter wave photons.
We assume that $T_E$ includes both the physical temperature variations 
and density variations. The angular correlation function for $T_B$ can then be
 written as
$$
C(\theta)\equiv \langle T_B({\boldsymbol \theta})\cdot T_B(0)\rangle 
=\int_{h/\sin\epsilon}^{(h+\Delta h)/\sin\epsilon}\!\!\!dz
\int_{h/\sin\epsilon}^{(h+\Delta h)/\sin\epsilon}\!\!\!dz'\kappa^2
\langle T_E(x,y,z')T_E(0,0,z)\rangle
$$
$$
=
\int_{h/\sin\epsilon}^{(h+\Delta h)/\sin\epsilon}\!\!\!dz
\int_{h/\sin\epsilon}^{(h+\Delta h)/\sin\epsilon}\!\!\!dz'\kappa^2C_{3D}(x,y,z'-z)\;.
$$
Fig.~\ref{fig:coord} shows the coordinate system used in this calculation. 
In the small angle regime,
the angular power spectrum is the Fourier transform of the angular correlation function, 
which can be derived from the 3-D correlation function, which in turn is the 3-D 
Fourier transform of the 3-D PSD:
$$
P(\alpha_x,\alpha_y)=
\int\!\!\int d\theta_x d\theta_y e^{-2\pi i (\alpha_x\theta_x+\alpha_y\theta_y)}
C(\theta)
$$
$$
=\int\!\!\int d\theta_x d\theta_y \int_{h/\sin\epsilon}^{(h+\Delta h)/\sin\epsilon} \!\!\!\!dz
\int_{h/\sin\epsilon}^{(h+\Delta h)/\sin\epsilon} \!\!\!\!dz' \kappa^2
e^{-2\pi i (\alpha_x\theta_x+\alpha_y\theta_y)}C_{3D}(x,y,z'-z)
$$
$$
=\int\!\!\int d\theta_x d\theta_y 
\int_{h/\sin\epsilon}^{(h+\Delta h)/\sin\epsilon} \!\!\!\!dz
\int_{h/\sin\epsilon}^{(h+\Delta h)/\sin\epsilon} \!\!\!\!dz' \int \!\!d^3k' \kappa^2
\cdot e^{-2\pi i (\alpha_x\theta_x+\alpha_y\theta_y)}
\cdot e^{2\pi i {\bf k}'\cdot \Delta {\bf x}} P_{3D}(k')\;,
$$
with $\alpha_i\theta_i=k_ix_i$ . 
Carrying out the all the integrals except for $dk'_z$, we have
$$
P(\alpha_x,\alpha_y)=\left(\frac{h}{\kappa\sin\epsilon}\right)^{-2}
\int_{-\infty}^{\infty}dk'_z P_{3D}(k_x,k_y,k'_z) \frac{\Delta h^2}{\sin^2\epsilon}
\left[\frac{\sin^2 (\pi k'_z \Delta h/\sin\epsilon) }{\pi^2k'^2_z \Delta h^2/\sin^2\epsilon}
\right]
$$
\begin{equation}
\approx \left(\frac{h}{\kappa\sin\epsilon}\right)^{-2}
\int_{-\sin\epsilon/2\Delta h}^{\sin\epsilon/2\Delta h} 
dk'_z P_{3D}(k_x,k_y,k'_z)\frac{\Delta h^2}{\sin^2\epsilon}
\approx P_{3D}
\left(\frac{\alpha_x}{h/\sin\epsilon},\frac{\alpha_y}{h/\sin\epsilon},0\right)
\frac{\Delta h}{\sin\epsilon}\left(\frac{h}{\kappa
\sin\epsilon}\right)^{-2}.\label{p2d}
\end{equation}
In the last two steps, the stationary phase approximation is used.
Eq.~[\ref{p2d}] states that the projected power spectrum 
obeys the same 3-D KT power law, $k^{-11/3}$, and its 
RMS is proportional to the square root of the path length through 
the turbulent layer, as would be expected.

With $P_{3D}\propto k^{-11/3}$, eq.~[\ref{p2d}] implies that 
\begin{equation}
P(\alpha)=B^2_{\nu}\sin(\epsilon)^{-8/3}(\alpha_x^2+\alpha_y^2)^{-11/6}.
\end{equation}
Comparing this expression with eq.[13] in LH2000, we can see that;
$$B^2_{\nu} = A h_{av}^{5/3}\,,$$
and has units of ${\rm mK}^2 \;{\rm rad}^{-5/3}$. 

\bibliographystyle{apj}

\bibliography{ms}

\end{document}